\documentclass[12pt]{article}
\usepackage{amssymb,latexsym, amsmath,color}
\usepackage{graphicx}
\usepackage{epsfig, picinpar,framed}
\usepackage[margin=2cm]{geometry}

\newcommand{\bfi}{\bfseries\itshape}

\newcommand{\rem}[1]{}
\newcommand{\url}[1]{}

\def\b{\begin{eqnarray}}
\def\e{\end{eqnarray}}
\def\n{\noindent}
\def\tfrac#1{\frac{ #1}}
\begin{document}

%%%%%%%%%%%%%%%%%%%%%%%%%%%%%%%%%
%%%%%%%%

\title{Smooth and Peaked Solitons of the CH equation}
\author{Darryl D. Holm$^{1}$  and Rossen I. Ivanov$^{1,2}$}
\addtocounter{footnote}{1}
\footnotetext{Department of Mathematics, Imperial College London. London SW7 2AZ, UK.
\texttt{d.holm@imperial.ac.uk, r.ivanov@imperial.ac.uk}
\addtocounter{footnote}{1}}
\footnotetext{School of Mathematical Sciences, Dublin Institute of Technology, Kevin Street, Dublin 8, Ireland,  \texttt{rivanov@dit.ie}
}

\date{}
\maketitle

\makeatother

\maketitle

%\rem{
%}

%%%%%%%%%%%%%%%%%%%%%%%%%%%%%%%%%%%%%%%%%%%%%%%%%%%%%%%%%%%%%%%%%%%%%%%%

%\doublespace

\begin{abstract}

The relations between smooth and peaked soliton solutions are reviewed for the Camassa-Holm (CH) shallow water wave equation in one spatial dimension. The canonical Hamiltonian formulation of the CH equation in action-angle variables  is expressed for solitons by using the scattering data for its associated  isospectral eigenvalue problem, rephrased as a Riemann-Hilbert problem. 
The momentum map from the action-angle scattering variables $T^*({\mathbb{T}^N})$ to the flow momentum ($\mathfrak{X}^*$) provides the Eulerian representation of the $N$-soliton solution of CH in terms of the scattering data and squared eigenfunctions of its  isospectral eigenvalue problem. 
The dispersionless limit of the CH equation and its resulting peakon solutions are examined by using an asymptotic expansion in the dispersion parameter. 
The peakon solutions of the dispersionless CH equation in one dimension are shown to generalize in higher dimensions to peakon wave-front solutions of the EPDiff equation whose associated momentum is supported on smoothly embedded subspaces. 
The Eulerian representations of the singular solutions of both CH and EPDiff are given by the (cotangent-lift) momentum maps arising from the left action of the diffeomorphisms on smoothly embedded subspaces.

\end{abstract}

\newpage

\tableofcontents

\section{Shallow water background for the CH equation}

Euler's equations for irrotational incompressible ideal fluid motion under
gravity with a free surface have an asymptotic expansion for shallow water
waves that contains two small parameters, $\epsilon$ and $\delta^2$, with
ordering $\epsilon\ge\delta^2$. These small parameters are
$\epsilon=a/h_0$ (the ratio of wave amplitude to mean depth) and
$\delta^2=(h_0/l_x)^2$ (the squared ratio of mean depth to horizontal length,
or wavelength). 
Euler's equations are made non-dimensional by introducing $x=l_x
x'$ for horizontal position, $z=h_0 z'$  for vertical position, $t=(l_x/c_0)
t'$ for time, $\eta=a \eta\,'$ for surface elevation and $\varphi=(g l_x
a/c_0)\varphi\,'$ for velocity potential, where $c_0=\sqrt{g h_0}$ is the mean
wave speed and $g$ is the constant gravity. The quantity $\sigma = \sigma\,' /
(h_0\rho c_0^2)$ is the dimensionless Bond number, in which $\rho$ is the mass
density of the fluid and  $\sigma'$ is its surface tension, both of which are
taken to be constants. After dropping primes, this asymptotic expansion yields
the nondimensional Korteweg-de Vries (KdV) equation for the horizontal velocity
variable $u=\varphi_x(x,t)$ at {\it linear} order in the small
dimensionless ratios $\epsilon$ and $\delta^2$, as the left hand side of
\begin{equation}\label{kdv-eqn}
u_t+u_x+\frac{3\epsilon}{2}uu_x
+\frac{\delta^2}{6}(1-3\sigma)u_{xxx}
=O(\epsilon\delta^2)
\,.
\end{equation}
Here, partial derivatives are denoted using subscripts, and boundary
conditions are $u=0$ and $u_x=0$ at spatial infinity on the real line. The
famous  $sech^2(x-ct)$ traveling wave solutions (the solitons) for KdV
(\ref{kdv-eqn}) arise in a balance between its (weakly) nonlinear steepening
and its third-order linear dispersion, when the quadratic terms in $\epsilon$
and $\delta^2$ on its right hand side are neglected.

On the right hand side of equation (\ref{kdv-eqn}), a normal form transformation due to Kodama
\cite{Ko1985} has been used to remove the other possible quadratic terms of
order $O(\epsilon^2)$ and $O(\delta^4)$. The remaining quadratic correction
terms in the KdV equation (\ref{kdv-eqn}) may be collected at order
$O(\epsilon\delta^2)$. These terms may be expressed, after introducing a
``momentum variable,'' 
\begin{equation}\label{mom-def}
m=u-\nu\delta^2u_{xx}
\,,
\end{equation}
and neglecting terms of {\it cubic} order in $\epsilon$ and $\delta^2$, as
\begin{equation}\label{b-eqn}
m_t+m_x+\frac{\epsilon}{2}(um_x+b\,mu_x)
+\frac{\delta^2}{6}(1-3\sigma)u_{xxx}
=0
\,.
\end{equation}
In the momentum variable $m=u-\nu\delta^2u_{xx}$, the parameter $\nu$ is given
by \cite{DuGoHo2001}
\begin{equation}\label{nu-def}
\nu=\frac{19-30\sigma-45\sigma^2}{60(1-3\sigma)}
\,.
\end{equation}
Thus, the effects of $\delta^2-$dispersion also enter the nonlinear terms.
After restoring dimensions in equation (\ref{b-eqn}) and rescaling velocity 
$u$ by $(b+1)$, the following ``$b$-equation" emerges,
\begin{equation}\label{dim-b-eqn}
m_t+c_0m_x
+um_x+b\,mu_x
+\Gamma u_{xxx}
=0
\,,
\end{equation}
where $m=u-\alpha^2u_{xx}$ is the dimensional momentum variable, and the
constants $\alpha^2$ and $\Gamma/c_0$ are squares of length scales.
When $\alpha^2\to0$, one recovers KdV from the $b$-equation (\ref{dim-b-eqn}),
up to a rescaling of velocity. Any value of the parameter $b\ne -1$ may be
achieved in equation (\ref{dim-b-eqn}) by an appropriate Kodama transformation
\cite{DuGoHo2001}. 

As we have emphasized, the values of the coefficients in the asymptotic
analysis of shallow water waves at quadratic order in their two small
parameters only hold, modulo the Kodama normal-form transformations. Hence,
these transformations may be used to advance the analysis and thereby gain
insight, by optimizing the choices of these coefficients. The freedom
introduced by the Kodama transformations among asymptotically equivalent
equations at quadratic order in $\epsilon$ and $\delta^2$ also helps to answer
the perennial question, ``Why are integrable equations so ubiquitous when one
uses asymptotics in modeling?'' Namely, there may be special values of the free parameters in the normal-form transformations of the asymptotics that allow integrability. 

\paragraph{Integrable cases of the $b$-equation (\ref{dim-b-eqn}).}
The cases $b=2$ and $b=3$ are special values. For these values, the $b$-equation
becomes completely integrable as a Hamiltonian system. For $b=2$, equation
(\ref{dim-b-eqn}) specializes to the integrable CH equation of Camassa and Holm
\cite{CaHo1993}. The case $b=3$ in (\ref{dim-b-eqn}) recovers the integrable 
DP equation of Degasperis and Procesi \cite{DePr1999}. These two cases exhaust
the integrable candidates for (\ref{dim-b-eqn}), as was shown using Painlev\'e
analysis. The $b-$family of equations (\ref{dim-b-eqn})
was also shown in \cite{MiNo2002} to admit the symmetry conditions necessary
for integrability, only in the cases $b = 2$ for CH and $b = 3$ for DP. 

The $b$-equation (\ref{dim-b-eqn}) with $b=2$ was first derived in 
Camassa and Holm \cite{CaHo1993} by using asymptotic expansions directly in
the Hamiltonian for Euler's equations governing inviscid incompressible flow
in the shallow water regime. In this analysis, the CH equation was shown to
be bi-Hamiltonian and thereby was found to be completely integrable by the
inverse scattering transform (IST) on the real line. This development of IST for CH equation (\ref{dim-b-eqn}) with $b=2$ is discussed in Section \ref{IST-sec}. 

Camassa and Holm \cite{CaHo1993} also discovered the remarkable peaked
soliton (peakon) solutions of (\ref{singlepeakon-soln},\ref{peakontrain-soln})
for the CH equation on the real line, given by (\ref{dim-b-eqn}) in the case
$b=2$. The peakons arise as solutions of (\ref{dim-b-eqn}), when $c_0 = 0$ and
$\Gamma = 0$ in the absence of linear dispersion.  Peakons move at a speed
equal to their maximum height, at which they have a sharp peak (jump in
derivative). The single peakon solution is
\begin{equation}\label{singlepeakon-soln-intro}
u(x,t)=ce^{-|x-ct|/\alpha}
\,,
\end{equation}
Unlike the KdV soliton, the peakon speed is independent of its
width ($\alpha$).  Periodic peakon solutions of CH were treated in Alber {\it
et al.} \cite{AlCaFeHoMa1999+2001}. There, the sharp peaks of
periodic peakons were associated with billiards reflecting at  the boundary of
an elliptical domain. These billiard solutions for the periodic peakons arise
from  geodesic motion on a tri-axial ellipsoid, in the limit that one of its
axes shrinks to zero length. 

The CH equation was found after its derivation as a shallow water equation in  \cite{CaHo1993} to fit into a class of integrable equations derived previously by using hereditary symmetries in Fokas and Fuchssteiner \cite{FoFu1981}.  See Fuchssteiner \cite{Fu1996} for an insightful history of how the shallow water equation (\ref{dim-b-eqn}) in the integrable case with $b=2$ relates to the mathematical theory of hereditary symmetries. 

Equation (\ref{dim-b-eqn}) with $b=2$ was recently re-derived as a shallow
water equation by using asymptotic methods in three different approaches in
Dullin {\it et al.} \cite{DuGoHo2001}, in Fokas and Liu \cite{FoLi1996} and
also in Johnson \cite{Jo2002}. These three derivations all used different
variants of the method of asymptotic expansions for shallow water waves in the
absence of surface tension. Only the derivation in Dullin {\it et al.} 
\cite{DuGoHo2001} took advantage of the parametric freedom in the Kodama normal-form transformations of the asymptotic expansion results at quadratic order.

The effects of the parameter $b$ on the solutions of equation (\ref{dim-b-eqn})
were investigated in Holm and Staley \cite{HoSt2003a}, where $b$
was treated as a bifurcation parameter, in the limiting case when the linear
dispersion coefficients are set to $c_0 = 0$ and $\Gamma = 0$. This
limiting case allows a variety of special solutions for different ranges of the values of $b$, in which the two nonlinear terms in equation (\ref{dim-b-eqn}) balance each other in the {\it absence} of linear dispersion.

Since its first appearance in \cite{CaHo1993}, the CH equation has been the centre of a confluence of scientific endeavors that includes water waves, integrable systems, PDE analysis, asymptotics, geometry and Lie groups. This confluence has led to a continued interest and opportunities for contributions from many different fields in mathematics. The interest in CH solutions may be measured by noticing that the entry `peakon' has acquired many thousands of  \emph{Google hits}.

\subsection*{Plan of the paper}
This paper aims to review some of the geometric highlights of recent work on the CH equation. It is certainly not exhaustive. It mainly focuses on comparing the soliton theory for smooth CH solutions with the peakon theory for its singular solutions in the dispersionless limit.
Section \ref{IST-sec} briefly explains the application of the method
of the inverse scattering transform (IST) for obtaining the soliton solutions of
CH equation. The set of scattering data is introduced and the formulation
of the inverse scattering as a Riemann-Hilbert problem is outlined. The solution
is expressed via the scattering data in a form that admits the peakon
limit in the sections that follow. 
In Section \ref{mommap-sec} the map between the action-angle variables (expressed
via the scattering data) and the momentum of the CH solution is formulated as a momentum map from the symplectic action-angle variables to the dual of the Lie algebra of smooth vector fields on the real line. This is a Poisson map, but the noncanonical bi-Hamiltonian structure of the CH equation which led to the discovery of its isospectral problem in \cite{CaHo1993} and its geometrical significance as geodesic flow on the diffeomorphism group are not discussed here.
In Section \ref{peakon-sec} we introduce the peakons as singular solutions
that appear in dispersionless limit. The $N$-peakon solution is governed by a finite-dimensional
integrable dynamical system.
Section \ref{peakon-limit-sec} presents the multi-peakon solution as a limiting case of the CH multi-soliton solution and points out the similarity between the dynamics of the peakon system and the well known Toda lattice. 
In Section \ref{superint-sec} we comment briefly on the existence of additional integrals
of motion of the peakon system, a property known as {\it superintegrability}.
The two-peakon system is analysed explicitly.
Section \ref{other-sing-sol-sec} presents the compacton and pulson solutions,
which are singular solutions, similar to peakons that can also be represented by the 
singular momentum map for peakons discussed in \cite{HoMa2004}. 
In Section \ref{EP-higher-sec} we deal with generalizations of the CH equation
in higher dimensions. These also allow for singular solutions. Although such
multidimensional generalizations are unlikely to be integrable, numerical studies
show that their solutions are stable and interact elastically in collisions. 
In Section \ref{Conclus-sec} we comment briefly on some limitations of our present discussion and point out three open problems for further research.

\section{Soliton Solutions of CH equation from Inverse Scattering}\label{IST-sec}

\subsection{Inverse scattering for the KdV equation}

One of the most significant results in the theory of nonlinear partial differential
equations was the development of Gardner, Greene, Kruskal and Miura \cite{GGKM1967,GGKM1974}
of a method for the exact solution of the initial-value problem for the KdV
equation. Prior to their work the only known exact solutions of KdV were
the travelling wave solutions. The idea is based on a representation of the
equation ($u_t+6uu_x+u_{xxx}=0$) as a compatibility condition of two eigenvalue problems with a time-independent
spectral parameter $\lambda$:
\b\label{KdV-isospec} 
\Psi_{xx}&+& u\Psi=\lambda \Psi \\
\Psi_t&=&(u_x+\gamma)\Psi-(4\lambda+2u)\Psi_x
\e

\n where $\gamma$ is an arbitrary constant. The method is conceptually analogous
in many ways to the Fourier transform method for solving linear equations.
This method is now known as Inverse Scattering Transform (IST). It was recast by
Lax \cite{Lax68} in a general framework that allows the IST to be used for
solving other nonlinear PDEs. An important consequence is the relationship between the discrete eigenvalues and the characteristics
of the solitons that emerge asymptotically. Another feature of soliton collisions
is the preservation of soliton identities after the interaction (asymptotically in time). For example, collisions of KdV solitons only result in a phase shift from the positions
where they would have been without the interaction.

The KdV was formulated as a completely integrable Hamiltonian system in a work
by Faddeev and Zakharov \cite{ZF1971}, the Hamiltonian form was also noted
by Gardner \cite{G1971}.    

Reviews of IST may be
found, for example, in Ablowitz et al. \cite{AbCl1991}, Dubrovin {\it et al.}
\cite{DuNoKr1985}, Novikov {\it et al.}
\cite{NoMaPiZa1984}. For discussions of other related bi-Hamiltonian
equations, see \cite{DePr1999}.

\subsection{Inverse scattering for CH solitons with dispersion}

In this section we outline the application of the IST for the CH equation. We use the form  with a linear dispersion term
\begin{equation}\label{eq1}
 u_{t}-u_{xxt}+2\omega u_{x}+3uu_{x}-2u_{x}u_{xx}-uu_{xxx}=0,
\end{equation}
where $\omega$ is a real constant. The equation in this form also appears
as a model of unidirectional propagation of shallow water waves over a flat
bottom \cite{CaHo1993, Jo2002, J03a} as well as that of axially
symmetric waves in a hyperelastic rod \cite{Dai98}. It can be obtained from
the $b$-equation (\ref{dim-b-eqn}) with $b=2$ via Galilean transformation
that removes the $u_{xxx}$ term.

Equation (\ref{eq1}) admits a Lax pair formulation \cite{CaHo1993} 
\b
\label{eq3} \Psi_{xx}&=&\frac{1}{4}\Big(1-(m+\omega)\Big)\Psi
 \\\label{eq4}
\Psi_{t}&=&-\left(\frac{2}{\lambda}+u\right)\Psi_{x}+\frac{u_{x}}{2}\Psi+\gamma\Psi
\e

\n where $\gamma$ is an arbitrary constant.  We will use this
freedom for a proper normalization of the eigenfunctions.

%%%%%%%%%%%%%%%%%%%%%%%%%%%%%%%%%%%%%%%%%%%%%%%%%%%%%%%%%%%%%%%%%%%%%%%%%%%%

\begin{figure}[ht]
\begin{center}
\includegraphics[scale=0.5,angle=-90]{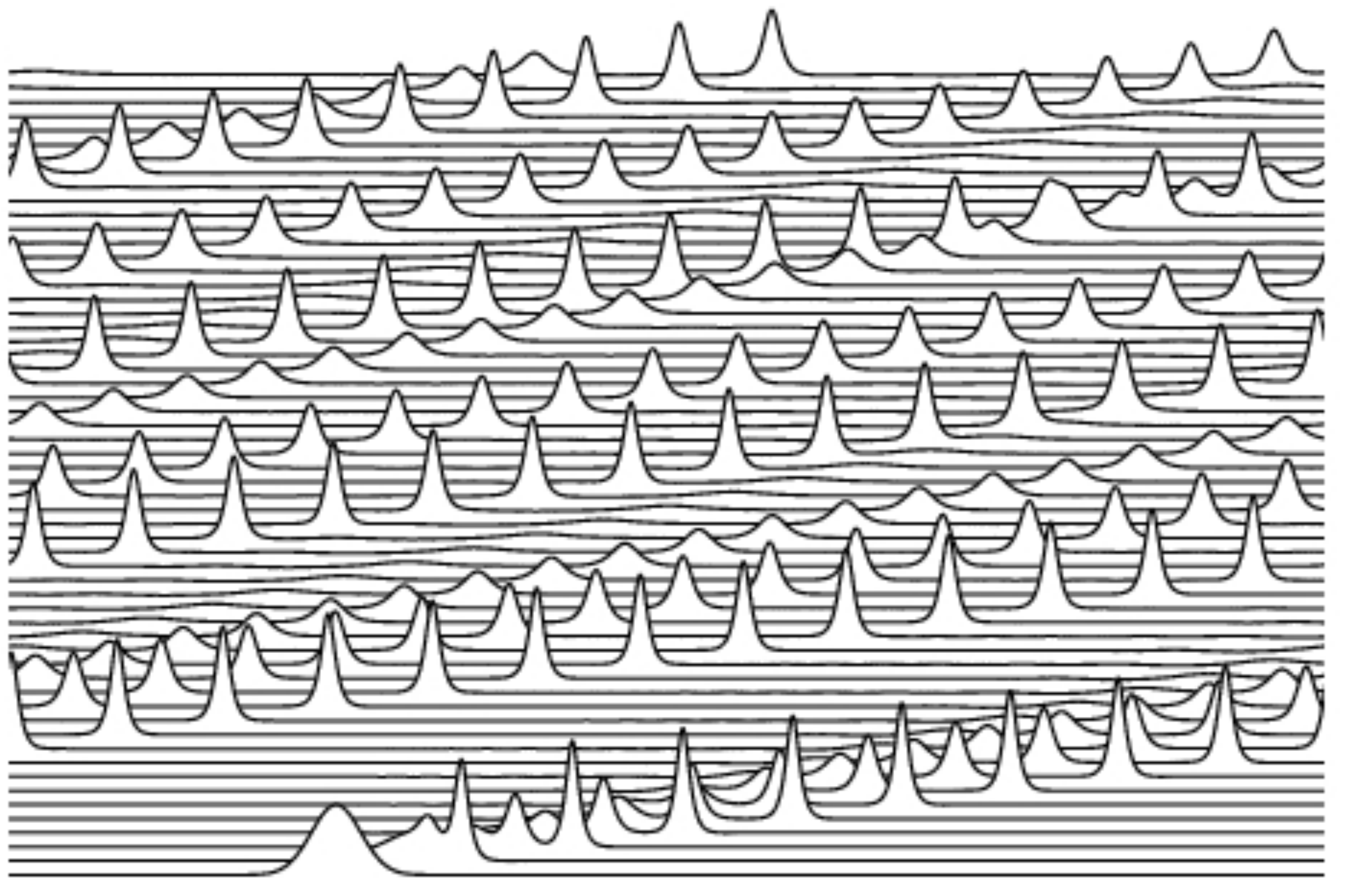}
\end{center}
\caption{
A Gaussian initial condition for the CH equation breaks
up into an ordered train of solitons as time evolves (the time direction
being vertical). The soliton train eventually wraps around the periodic
domain, thereby allowing the leading solitons to overtake the slower
emergent solitons from behind in collisions that cause phase shifts, as
discussed in \cite{CaHo1993}.}
\label{soliton_figure}
\end{figure}

%%%%%%%%%%%%%%%%%%%%%%%%%%%%%%%%%%%%%%%%%%%%%%%%%%%%%%%%%%%%%%%%%%%%%%%%%%%%

In our further considerations $m$ will be a Schwartz class
function, $\omega >0$ and $m(x,0)+\omega > 0$. Then $m(x,t)+\omega
> 0$ for all $t$ \cite{C05}. Let us introduce a new spectral parameter $k$,
such that \b \label{lambda}
\lambda(k)= \frac{1}{\omega}\Big(1+ 4k^{2}\Big).\e

The spectrum of the problem (\ref{eq3}) under these conditions is
described in \cite{C01}. The continuous spectrum in terms of $k$
corresponds to $k$ -- real. The discrete spectrum (in the upper
half plane) consists of finitely many points $k_{n}=i\kappa _{n}$,
$n=1,\ldots,N$ where $\kappa_{n}$ is real and $0<\kappa_{n}<1/2$.

For all real $k\neq 0$ a basis in the space of solutions of
(\ref{eq3}) can be introduced, $\psi(x,k)$ and
$\bar{\psi}(x,\bar{k})$, fixed by its asymptotic when
$x\rightarrow\infty$  \cite{C01}: \b \label{eq5}
\psi(x,k)&=&e^{-ikx}+o(1), \qquad x\rightarrow\infty \e

\n   Another basis can be introduced, $ \varphi(x,k)$ and $
\bar{\varphi}(x,\bar{k})$ fixed by its asymptotic when
$x\rightarrow -\infty$: \b \label{eq5a} \varphi
(x,k)&=&e^{-ikx}+o(1), \qquad x\rightarrow -\infty \e
and the relation between the two bases is \cite{CGI06} \b
\label{eq8} \varphi(x,k)=a(k)\psi(x,k)+b(k)\bar{\psi}(x,k). \e

\n where
\b \label{eq10} |a(k)|^{2}-|b(k)|^{2}=1. \e

The quantity $\mathcal{R}(k)=b(k)/a(k)$ is called the reflection
coefficient, without any danger of confusing $b(k)$ with the bifurcation parameter $b$ introduced in (\ref{dim-b-eqn}). 

\n All of the required information about the scattering, i.e. $a(k)$ and
$b(k)$, is provided by $\mathcal{R}(k)$ for $k>0$ only \cite{CI06}.
It is sufficient to know $\mathcal{R}(k)$ only on the half line
$k>0$, since $\bar{a}(k)=a(-k)$, $\bar{b}(k)=b(-k)$ and thus
$\mathcal{R}(-k)=\bar{\mathcal{R}}(k)$.

The constant $\gamma$ in (\ref{eq4}) can be chosen for each eigenfunction
in such a way that $a(k)$ does not depend on $t$ and is a generating function
of the integrals of motion \cite{CI06}. At the points of the discrete spectrum, $a(k)$ has simple zeroes
\cite{C01}, therefore  $\varphi$ and $\bar{\psi}$ are linearly
dependent (\ref{eq8}): 
\begin{equation} \label{eq200}
\varphi(x,i\kappa_n)=b_n\bar{\psi}(x,-i\kappa_n).
\end{equation}

\n  In other words, the discrete spectrum is simple, there is only
one (real) eigenfunction $\varphi^{(n)}(x)$, corresponding to each
eigenvalue $i\kappa_n$, and we can take this eigenfunction to be
\b \label{eq201}\varphi^{(n)}(x)\equiv \varphi(x,i\kappa_n)\e

\n The asymptotic behaviour of $\varphi^{(n)}$, according to (\ref{eq5a}),
(\ref{eq200}) is  \b \label{eq203} \varphi^{(n)}(x)&=&e^{\kappa_n
x}+o(e^{\kappa_n x}), \qquad x\rightarrow -\infty;
\\ \label{eq204}
\varphi^{(n)}(x)&=& b_n e^{-\kappa_n x}+o(e^{-\kappa_n x}), \qquad
x\rightarrow \infty. \e

\n The sign of $b_n$ obviously depends on the number of the zeroes
of $\varphi^{(n)}$. Suppose that \b \label{eqN}
0<\kappa_{1}<\kappa_{2}<\ldots<\kappa_{N}<1/2.\e Then from the
oscillation theorem for the Sturm-Liouville problem \cite{B},
$\varphi^{(n)}$ has exactly $n-1$ zeroes. Therefore \b
\label{eq205} b_n= (-1)^{n-1}|b_n|.\e

The set \b \label{eq206} \mathcal{S}\equiv\{ \mathcal{R}(k)\quad
(k>0),\quad \kappa_n,\quad R_n=b_n/ia'(i\kappa_n),\quad n=1,\ldots
N\} \e

\n is called the \emph{scattering data}. The Hamiltonians for the CH equation
in terms of the scattering data are presented in \cite{CI06}. The
time evolution of the scattering data is \cite{CGI06}:

\b \label{eq19} \mathcal{R}(k,t)&=&\mathcal{R}(k,0)e^{-\frac{4i
k}{\lambda }t}. \\
\label{eq207} R_n(t)&=&R_n(0)\exp\left(\frac{4\kappa_n}{\lambda_n}t\right), \e where $R_n(t)$ is always a
positive quantity \cite{CGI06} and $\lambda_n=\lambda(\kappa_n)$.

The scattering coefficient $a(k)$ is analytic for
$\mathrm{Im}\phantom{*} k>0$ with asymptotic \cite{CI06}
\begin{equation}\label{eqi9}
 e^{i\beta k}a(k)\rightarrow 1,
 \qquad |k|\rightarrow\infty.
\end{equation}

\n where $\beta$ is a constant (integral of motion):
\begin{equation}\label{eqi8}
 \beta= \int
 _{-\infty}^{\infty}\Big(\sqrt{1+\frac{m(x)}{\omega}}-1\Big)dx.
\end{equation}

The asymptotics of the Jost solutions (or, rather the following
quantities, depending on the Jost solutions) for $|k|\rightarrow
\infty$ have the form \cite{CGI06}

\b\label{eq20} \underline{\psi}(x,k)&\equiv&
\psi(x,k)e^{\frac{iky}{\sqrt{\omega}}}=X_0(x)+\frac{X_1(x)}{k}+\frac{X_2(x)}{k^2}+\ldots,
\\
\underline{\varphi}(x,k) &\equiv&
\varphi(x,k)e^{ik(\frac{y}{\sqrt{\omega}} +
\beta)}=X_0(x)+\frac{\widetilde{X}_1(x)}{k}+\frac{\widetilde{X}_2(x)}{k^2}+\ldots
\label{eq21}\e

\n where $X_0(x)=[\omega/(m(x)+\omega)]^{1/4}>0$, \b\label{eq22}
y(x)=\kappa\left[x+\int_{\infty}^x(X_0^{-2}(\tilde{x})-1)d\tilde{x}\right].\e

\n Moreover, (\ref{eq20}) is analytic for $\mathrm{Im}\phantom{*}
k<0$, (\ref{eq21}) is analytic for $\mathrm{Im}\phantom{*} k>0$
\cite{CGI06}.

From (\ref{eq8}) with (\ref{eq20}) and (\ref{eq21}) we obtain

\b \label{eq27}
\frac{\underline{\varphi}(x,k)}{e^{ik\beta}a(k)}=\underline{\psi}(x,k)+
\mathcal{R}(k)\bar{\underline{\psi}}(x,k)e^{2iky(x)/\sqrt{\omega}}. \e

\n The function $\underline{\varphi}(x,k)/(e^{ik\beta}a(k))$ is
analytic for $\mathrm{Im}\phantom{*} k>0$, $\underline{\psi}(x,k)$
is analytic for $\mathrm{Im}\phantom{*} k<0$. Thus, (\ref{eq27})
represents an additive {\bfi Riemann-Hilbert Problem (RHP)} with a jump
on the real line, given by
$\mathcal{R}(k)\bar{\underline{\psi}}(x,k)e^{2iky(x)/\sqrt{\omega}}$.

The solution of this RHP follows a standard pattern, see
\cite{CGI06} for more details. The solution simplifies considerably in
the case $\mathcal{R}(k)\equiv 0$. This is the $N$-soliton
solution:

\b\label{solk-} \underline{\psi}(x,k)&=&X_0(x)-i\sum_{n=1} ^N
\frac{R_n(t)e^{-2\kappa_ny/\sqrt{\omega}}\underline{\psi}(x,-i\kappa_n)}{i\kappa_n-k},
\quad \mathrm{Im}\phantom{*} k<0,
\\
\frac{\underline{\varphi}(x,k)}{e^{ik\beta}a(k)}&=&X_0(x)-i\sum_{n=1}
^N
\frac{R_n(t)e^{-2\kappa_ny/\sqrt{\omega}}\underline{\psi}(x,-i\kappa_n)}{i\kappa_n-k},
\quad \mathrm{Im}\phantom{*} k>0. \label{solk+} \e

\n From (\ref{solk-}) one has a linear system for the quantities
$\underline{\psi}(x,-i\kappa_n,t)$ with a solution

\b \label{eq43a}
\underline{\psi}(x,-i\kappa_n,t)=X_0(x)\sum_{p=1}^N
A^{-1}_{np}[y,t], \quad n=1,\ldots,N.\e

\n where

\b \label{eq43} A_{pn}[y,t]\equiv
\delta_{pn}+\frac{R_n(t)e^{-2\kappa_ny/\sqrt{\omega}}}{\kappa_p+\kappa_n}.
\e

Taking $k=-i/2$ in (\ref{solk-}) produces

\b e^{-\frac{1}{2}(x-\frac{y}{\sqrt{\omega}})}&\equiv&
\underline{\psi}(x,-i/2)=X_0(x)\left(1-\sum_{n,p=1} ^N
\frac{R_n(t)e^{-2\kappa_ny/\sqrt{\omega}}}{\kappa_n+\frac{1}{2}}A^{-1}_{np}[y,t]\right).
\nonumber \\
\label{-i/2} \e

The substitution $\kappa=i/2$ in (\ref{solk+}) with $a(i/2)=1$
gives

\b e^{\frac{1}{2}(x-\frac{y}{\sqrt{\omega}})}&\equiv&
\frac{\underline{\varphi}(x,i/2)}{e^{-\beta/2}a(i/2)}=X_0(x)\left(1-\sum_{n,p=1}
^N
\frac{R_n(t)e^{-2\kappa_ny/\sqrt{\omega}}}{\kappa_n-\frac{1}{2}}A^{-1}_{np}[y,t]\right).
\nonumber \\
\label{i/2} \e

From (\ref{-i/2}) and (\ref{i/2}) there follows a parametric
representation

\b x&=&X(y,t)\equiv \frac{y}{\sqrt{\omega}}+\ln \frac{f_+}{f_-}, \label{X=f/f}\\
f_{\pm}& \equiv & 1-\sum_{n,p=1} ^N
\frac{R_n(t)e^{-2\kappa_ny/\sqrt{\omega}}}{\kappa_n\mp\frac{1}{2}}A^{-1}_{np}[y,t]
.
\label{f pm} \e

\subsection{Parametric form of the dispersive CH soliton solution}

From (\ref{eq22}) and (\ref{X=f/f}) one can compute the solution
in parametric form \b \label{eq u} u(X(y,t),t)=X_t(y,t),\qquad
x=X(y,t), \label{sol}\e

\n where $X(y,t)$ is given in terms of the scattering data in
(\ref{X=f/f}), (\ref{f pm}).

Upon introducing new notations  \b \label{xi_j} \xi_j&=&2\kappa_j\left(
-\frac{y}{\sqrt{\omega}}+\frac{2t}{\lambda_j}+x_{j0}\right) \\
x_{j0}&=&\frac{1}{2\kappa_j}\ln \frac{R_j(0)}{2\kappa_j} \label{x_0j}\\
\phi_j &=& \ln \frac{1-2\kappa_j}{1+2\kappa_j}\label{phi_j}\\
\gamma_{ij}&=& \ln
\left(\frac{\kappa_i-\kappa_j}{\kappa_i+\kappa_j}\right)^2
\label{gamma ij} \e

\n one can rewrite the expression for $f_{\pm}$ (\ref{f pm}) in
the form \cite{Ma05,PI,PII,PIII}

\b f_{\pm}& \equiv & \sum_{\mu =0,1}\exp
\left[\sum_{i=1}^{N}\mu_i(\xi_i\mp\phi_i)
+
\!\!\sum_{1\le i<j \le j}
\mu_i \mu_j \gamma_{i j}\right].\e

The solution for $m$ can be obtained from (\ref{sol}). First we notice  that
$$ \frac{\partial u(X(y,t),t)}{\partial y}= u_x(X(y,t),t)X_y. $$ and also

$$ \frac{\partial u(X(y,t),t)}{\partial y}= \frac{\partial X_t}{\partial y}=X_{t y}.$$ Thus $u_x(X(y,t),t)=X_{t y}/X_y$. Similarly, $$u_{xx}(X(y,t),t)=\frac{1}{X_{y}}\left
(\frac{X_{t y}}{X_y}\right)_y$$ and $$ m(X(t,y),t)=u(X(t,y),t)-u_{xx}(X(t,y),t)=X_t-\frac{1}{X_{y}}\left
(\frac{X_{t y}}{X_y}\right)_y;$$

\b 
m(x,t)&=&\int_{-\infty}^{\infty}P(y,t)\delta(x-X(y,t)) \text{d} y \label{m(x,t)}
\\
&\text{with}& \qquad P(y,t)=X_t X_y-\left(\frac{X_{t y}}{X_y} \right)_y  \label{P(y,t)}, \\
u(x,t)&=&\frac{1}{2}\int_{-\infty}^{\infty}P(y,t)\exp(-|x-X(y,t)|) \text{d} y.\label{u(x,t)}
\e
As we shall see in Section \ref{EP-higher-sec}, this representation is also useful in the study of multidimensional solutions.

\subsection{Relation to KdV hierarchy}\label{KdV-sec}

The spectral problem (\ref{eq3}) is gauge equivalent to a
standard Sturm-Liouville problem, well known from the KdV
hierarchy, cf. (\ref{KdV-isospec}), with short notation $q=m+\omega$: 
\b -\Phi_{yy}+U(y)\Phi&=&\mu \Phi, \qquad
\mu = \frac{\lambda}{4}-\frac{1}{4\omega}, \nonumber \\
\Phi(y)&=&q^{1/4}\Psi, \qquad \frac{dy}{dx}=\sqrt{q}\,,
\label{chvar} \\
U(y)&=&\frac{1}{4q(y)}+\frac{q_{yy}(y)}{4q(y)}-\frac{3q_y^2(y)}{16q^2(y)}-\frac{1}{4\omega}.
\label{EP} \e

\noindent
Note that (\ref{chvar}) leads to two possible expressions for the
change of the variables in the Liouville transformation: \b y&=&
\sqrt{\omega}x+\int_{-\infty}^x (\sqrt{q(x')}-\sqrt{\omega}){\text
d}x'+ \text{const}, \label{y1} \\ y&=&
\sqrt{\omega}x+\int_{\infty}^x (\sqrt{q(x')}-\sqrt{\omega}){\text
d}x'+ \text{const}. \label{y2} \e

\n These two possibilities, (\ref{y1}), (\ref{y2}) are only
consistent iff \b \int_{-\infty}^{\infty}
(\sqrt{q(x)}-\sqrt{\omega}){\text d}x=\text{constant}, \nonumber \e
which is always the case, since the integral under question is (up
to a multiplier) the Casimir function $\beta$ (\ref{eqi8}).

The matching of the CH hierarchy to KdV hierarchy requires solving
the Ermakov-Pinney equation (\ref{EP}) \cite{C01}, which is not
straightforward and leads to the same solution (\ref{sol}) in parametric form.

\section{Momentum map formulation with action-angle variables}\label{mommap-sec}

The canonical  Poisson brackets for the scattering data of the CH
equation are computed in \cite{CI06} where also the action-angle
variables are expressed in terms of the scattering data. Let us consider the action variable for the N-soliton solution. (These considerations can be extended easily to the variables of the continuous spectrum.) The angle variable is $\Phi_n = \ln R_n(t)$; it is linear in time $t$ and $\dot{\Phi}_n = 4\kappa_n/\lambda_n .$ Let us introduce $\Lambda_n := 4\kappa_n/\lambda_n$ into Hamilton's principle $\delta S=0$, with
\[
S[u,\Phi_n,\Pi_n]
=
\int\bigg( \ell[u]+\sum_{n=1}^{N}\Pi_n\Big(\dot{\Phi}_n -\Lambda_n [u] \Big) \bigg) dt.
\]
Here the Lagrange multiplier $\Pi_n$ enforces the action-angle relation for the CH scattering data as a $\mathbb{T}^N$ shift of the angles $\Phi_n$ at constant angular frequencies $\Lambda_n$, with $n=1,\dots,N$. The stationary variation is
\[
0 = \delta S
=
\int\left(\bigg(\frac{\delta \ell}{\delta u}
-
\sum_{n=1}^{N}\Pi_n \frac{\delta \Lambda_n}{\delta u} \bigg) \delta u
+ 
\sum_{n=1}^{N}\Big(\dot{\Phi}_n -\Lambda_n [u] \Big)\delta \Pi_n 
- \sum_{n=1}^{N}\dot{\Pi}_n \delta \Phi_n \right) dt.
\]
Since by definition $ m= \delta \ell /\delta u$ is the momentum, $\delta S=0$ implies  the Eulerian representation  
\b 
\label{mommap1}
m=\frac{\delta \ell}{\delta u}=\sum_{n=1}^{N}\Pi_n \frac{\delta \Lambda_n}{\delta u}
\,, \qquad \text{with}  \\
\dot{\Phi}_n = \Lambda_n [u]
\qquad \text{and} \qquad 
\dot{\Pi}_n=0. 
\label{toralNaction}
\e
Relation (\ref{mommap1}) is the \emph{momentum map} 
\[
(\Phi_n,\Pi_n)\in T^*\mathbb{T}^N\to m\in\mathfrak{X}^*
\]
for the toral $\mathbb{T}^N$ action (\ref{toralNaction}) on the angles $\Phi_n$ at constant angular frequencies $\Lambda_n$. This momentum map from the action-angle scattering variables $T^*({\mathbb{T}^N})$ to the flow momentum $\mathfrak{X}^*(\mathbb{R})$ (dual to the smooth vector fields $\mathfrak{X}(\mathbb{R})$ on the real line) provides the Eulerian representation of the $N$-soliton solution of CH in terms of the scattering data and squared eigenfunctions of its  isospectral eigenvalue problem. 
Momentum maps for Hamiltonian
dynamics are reviewed in \cite{MaRa1999}, for example.

By using the spectral quantities of the $N$-soliton solution (recall: 
$\lambda_n=(1-4\kappa_n^2)/\omega $)
one may express the variation of the spectrum with respect to the CH solution in terms of the squared-eigenfunctions of the isospectral problem as \cite{CI06} 
\[
\frac{\delta \Lambda_n
}{\delta m(x,t)}=\frac{(1+4\kappa_n^2)}{2\omega\kappa_n
\lambda_n}R_n(t)[\bar{\psi}(x,-i\kappa_n, t)]^2, 
\]
in which $\bar{\psi}(x,-i\kappa_n, t)$ is the eigenfunction that belongs to eigenvalue $\lambda_n$, see (\ref{eq200}). 

On the other hand, the expansion of $u(x,t)$ over squares of eigenfunctions is given by \cite{CGI07}
\[
u(x,t)= \sum_{n=1}^N \frac{4 \kappa_n}{\omega\lambda_n^2} R_n(t)[\bar{\psi}(x,-i\kappa_n, t)]^2
.
\]

\n Consequently, 

\[
m(x,t)= \sum_{n=1}^N \frac{4 \kappa_n}{\omega\lambda_n^2} R_n(t)(1-\partial^2)[\bar{\psi}(x,-i\kappa_n, t)]^2, 
\]
or  
\[
m(x,t)= \sum_{n=1}^N \Pi_n J_n(x,t)\,, 
\]
where $\Pi_n$ and $J_n(x,t)$ denote explicitly 

\b \Pi_n&=&
\frac{8 \kappa_n^2}{\lambda_n(1+4\kappa_n^2)}
=
\frac{2 \Lambda_n\kappa_n}{1+4\kappa_n^2}
\,, \\
J_n (x,t)&\equiv& \frac{\delta \Lambda_n
}{\delta u(x,t)}
\\
&=& \frac{(1+4\kappa_n^2)}{2\omega\kappa_n
\lambda_n} R_n(t)(1-\partial^2)[\bar{\psi}(x,-i\kappa_n, t)]^2
. 
\e 
Thus, the momentum map  (\ref{mommap1}) from the action-angle variables under going dynamics (\ref{toralNaction}) to the Eulerian representation of the momentum for the CH solution is expressed in terms of the scattering data and squared eigenfunctions of its $N$-soliton isospectral eigenvalue problem. Perhaps not unexpectedly, this momentum map may be applied to the action-angle representation of the solution of any integrable Hamiltonian PDE.

%%%%%%%%%%%%%%%%%%%%%%%%%%%%%%%%%%%%%%%%%%%%%%%%%%%%%%%%%%%%%%%%%%%%%%%%

\section{Peakons}  \label{peakon-sec}

\subsection{Peakons: the singular solution ansatz} 
Camassa and Holm \cite{CaHo1993} discovered the ``peakon'' solitary
traveling wave solution for a shallow water wave, 
\begin{equation}\label{singlepeakon-soln}
u(x,t)=ce^{-|x-ct|/\alpha}
\,,
\end{equation}
whose fluid velocity $u$ is a function of position $x$ on the real line and
time $t$.  The peakon traveling wave moves at a speed equal to its maximum
height, at which it has a sharp peak (jump in derivative).  Peakons are an
emergent phenomenon, solving the initial value problem for a partial
differential equation derived by an asymptotic expansion of Euler's equations
using the small parameters of shallow water dynamics. Peakons are {\it
nonanalytic} solitons, which superpose as 
\begin{equation}\label{peakontrain-soln}
u(x,t)=\frac12\sum_{a=1}^Np_a(t)e^{-|x-q_a(t)|/\alpha}
=: \frac12\sum_{a=1}^N p_a(t) g(x-q_a(t))/\alpha)
\,,
\end{equation}
for sets $\{p\}$ and $\{q\}$ satisfying canonical Hamiltonian dynamics. 
Peakons arise for shallow water waves in the limit of zero linear
dispersion in one dimension.  Peakons satisfy a partial differential equation
(PDE) arising from Hamilton's principle for geodesic motion on the smooth
invertible maps (diffeomorphisms) with respect to the $H^1$ Sobolev norm of
the fluid velocity. Peakons generalize to higher dimensions, as well. We
explain how peakons were derived in the context of shallow water asymptotics
and describe some of their remarkable mathematical properties. 

Peakons were first found as singular soliton solutions of the
completely integrable CH equation. This is equation (\ref{dim-b-eqn}) with
$b=2$, now rewritten in terms of the velocity, as
\begin{eqnarray}\label{CH-u-eqn}
u_t+c_0u_x+3uu_x&+&\Gamma u_{xxx}
\nonumber\\
&=&\alpha^2(u_{xxt}+2u_xu_{xx}+uu_{xxx})
\,.%\,,\nonumber\quad({\rm CH})
\end{eqnarray}
Peakons were found in \cite{CaHo1993} to arise in the absence of linear
dispersion. That is, they arise when $c_0=0$ and $\Gamma=0$ in CH
(\ref{CH-u-eqn}). Specifically, peakons are the individual terms in the peaked
$N-$soliton solution of CH (\ref{CH-u-eqn}) for its velocity,
in the absence of linear dispersion.
Each term in the sum (\ref{peakontrain-soln}) is a solition with a sharp peak at its maximum. Hence,
the name ``peakon.'' Expressed using its momentum,
$m=(1-\alpha^2\partial_x^2)u$, the peakon velocity solution
(\ref{peakontrain-soln}) of dispersionless CH becomes a sum over a delta
functions, supported on a set of points moving on the real line. Namely, 
the peakon velocity solution (\ref{peakontrain-soln}) implies
\begin{equation}\label{CHpeakon-m-soln}
m(x,t)=\alpha\sum_{a=1}^N \, p_a(t)\delta(x-q_a(t))
\,,\end{equation}
because of the relation
$(1-\alpha^2\partial_x^2)e^{-|x|/\alpha}=2\alpha\delta(x)$. These
solutions satisfy the $b$-equation (\ref{dim-b-eqn}) for any value of $b$,
provided $c_0 = 0$ and $\Gamma = 0$. As we shall discuss later, the peakon momentum relation (\ref{CHpeakon-m-soln}) is again a momentum map. 

Thus, peakons are {\it singular momentum solutions} of the dispersionless
$b$-equation, although they are not stable for every value of $b$. From
numerical simulations \cite{HoSt2003a}, peakons are conjectured to
be stable for $b>1$. In the integrable cases $b = 2$ for CH and $b = 3$ for DP,
peakons are stable singular {\it soliton} solutions. The spatial velocity
profile $e^{-|x|/\alpha}/(2\alpha)$ of each separate peakon in
(\ref{peakontrain-soln}) is the Green's function for the Helmholtz operator on
the real line, with vanishing boundary conditions at spatial infinity. Unlike
the KdV soliton, whose speed and width are related, the width of the peakon
profile is set by its Green's function, independently of its speed.

\subsection{Integrable peakon dynamics of CH} 
Substituting the peakon solution ansatz (\ref{peakontrain-soln})
and (\ref{CHpeakon-m-soln}) into the dispersionless CH equation,
\begin{equation}\label{DCHmom-eqn}
m_t
+
um_x
+
2mu_x
=0
\,,\quad\hbox{with}\quad
m=u-\alpha^2u_{xx}
\,,
\end{equation}
yields {\it Hamilton's canonical equations} for the dynamics of the discrete
set of peakon parameters $p_a(t)$ and $q_a(t)$, 
\begin{equation}\label{Ham-peakon-eqn}
\dot{q}_a(t) = \frac{\partial h_N}{\partial p_a}
\quad\hbox{and}\quad
\dot{p}_a(t) = -\,\frac{\partial h_N}{\partial q_a}
\,,
\end{equation}
for $a=1,2,\dots,N$, with Hamiltonian given by \cite{CaHo1993},
\begin{equation}\label{H-peakon-ansatz}
h_N=\tfrac{1}{4}\sum_{a,b=1}^N p_a\,p_b\,e^{-|q_a-q_b|/\alpha}
\,.
\end{equation}
Or explicitly, 
\b 
\dot{q}_a&=&\frac{1}{2}\sum_{b=1}^Np_b e^{-|q_a-q_b|/\alpha} \label{q_a}
\\
\dot{p}_a&=&\frac{p_a}{2\alpha}\sum_{b=1}^N p_b e^{-|q_a-q_b|/\alpha}\text{sgn}(q_a-q_b) \label{p_a}
\e

Thus, one finds that the points $x=q_a(t)$ in the peakon solution
(\ref{peakontrain-soln}) move with the flow of the fluid velocity $u$ at
those points, since $u(q_a(t),t)=\dot{q}_a(t)$. This means the $q_a(t)$ are
{\it Lagrangian} coordinates. Moreover, the singular momentum solution 
(\ref{CHpeakon-m-soln}) is the Lagrange-to-Euler map for an invariant
manifold of the dispersionless CH equation (\ref{DCHmom-eqn}). On this
finite-dimensional invariant manifold for the partial differential
equation (\ref{DCHmom-eqn}), the dynamics is canonically Hamiltonian. 

With Hamiltonian (\ref{H-peakon-ansatz}), the canonical
equations (\ref{Ham-peakon-eqn}) for the $2N$ canonically conjugate peakon
parameters $p_a(t)$ and $q_a(t)$ were interpreted in \cite{CaHo1993}
as describing {\it geodesic motion} on the
$N-$dimensional Riemannian manifold whose co-metric is
$g^{ab}(\{q\})=e^{-|q_a-q_b|/\alpha}$. Moreover, the canonical geodesic
equations arising from Hamiltonian (\ref{H-peakon-ansatz}) comprise an
integrable system for any number of peakons $N$. This integrable system was
studied in \cite{CaHo1993} for solutions on the real line, and in
\cite{AlCaFeHoMa1999+2001,McCo1999} and references therein, for spatially
periodic solutions.  \rem{The integrable solutions on the real line were
related to the Toda chain with open ends, in \cite{Va2003}.}

The integrals generated by the action variables in terms of the coordinates
can be recovered as tr$(L^k)$, where $L$ is the Lax operator for the peakon
system \cite{CaHo1993}.

Being a completely integrable Hamiltonian soliton equation, the continuum CH
equation (\ref{CH-u-eqn}) has an associated isospectral eigenvalue problem,
discovered in \cite{CaHo1993} for any values of its dispersion parameters
$c_0$ and $\Gamma$. Remarkably, when $c_0 = 0$ and $\Gamma = 0$, this
isospectral eigenvalue problem has a purely {\it discrete} spectrum. Moreover,
in this case, each discrete eigenvalue corresponds precisely to the
time-asymptotic velocity of a peakon. This discreteness of the CH isospectrum
in the absence of linear dispersion implies that {\it only} the singular
peakon solutions (\ref{CHpeakon-m-soln}) emerge asymptotically in time, in the
solution of the initial value problem for the dispersionless CH equation 
(\ref{DCHmom-eqn}). This is borne out in numerical simulations of the
dispersionless CH equation (\ref{DCHmom-eqn}), starting from a smooth initial
distribution of velocity \cite{FrHo2001,HoSt2003a}. 

\begin{figure}[ht]
\begin{center}
\includegraphics[scale=0.75,angle=0]{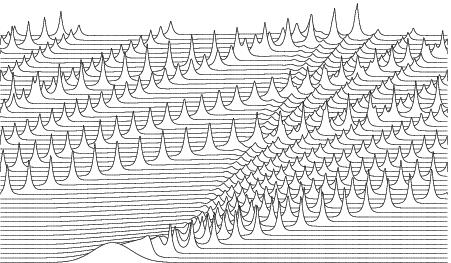}
\end{center}
\caption{%\footnotesize
A Gaussian initial condition for the CH equation breaks
up into an ordered train of peakons as time evolves (the time direction
being vertical). The peakon train eventually wraps around the periodic
domain, thereby allowing the leading peakons to overtake the slower
emergent peakons from behind in collisions that cause phase shifts, as
discussed in \cite{CaHo1993}.}
\label{peakon_figure}
\end{figure}

Figure \ref{peakon_figure}
shows the emergence of peakons from an initially Gaussian
velocity distribution and their subsequent elastic collisions in
a periodic one-dimensional domain. This
figure demonstrates that singular solutions dominate the initial
value problem and, thus, that it is imperative to go beyond
smooth solutions for the CH equation; the situation is similar
for the EPDiff equation.

%%%%%%%%%%%%%%%%%%%%%%%%%%%%%%%%%%%%%%%%%%%%%%%%%%%%%%%%%%%%%%%%%%%%%%%%%%%%

%%%%%%%%%%%%%%%%%%%%%%%%%%%%%%%%%%%%%%%%%%%%%%%%%%%%%%%%%%%%%%%%%%%%%%%%%%%%

\paragraph{Peakons as mechanical systems.} 
Governed by canonical Hamiltonian equations, each $N-$peakon solution
can be associated with a mechanical system of moving particles. Calogero et
al. \cite{Ca1995} further extended the class of mechanical systems of this
type. The r-matrix approach was applied to the Lax pair formulation of the
$N-$peakon system for CH by Ragnisco and Bruschi 
\cite{RaBr1996}, who also pointed out the connection of this system with the
classical Toda lattice. A discrete version of the Adler-Kostant-Symes
factorization method was used by Suris \cite{Su1996} to study a
discretization of the peakon lattice, realized as a discrete integrable system
on a certain Poisson submanifold of $gl(N)$ equipped with an r-matrix Poisson
bracket. Beals {\it et al.} \cite{BeSaSm1999+2000} used the
Stieltjes theorem on continued fractions and the classical moment problem for
studying multi-peakon solutions of the CH equation. Generalized peakon systems
are described for any simple Lie algebra by Alber {\it et al.}
\cite{AlCaFeHoMa1999+2001}.

\section{Peakon limit of the CH soliton solutions}\label{peakon-limit-sec}

The limit $\omega \to 0$ in the $N$-soliton solution $u(x,t)$ produces the $N$-peakon solution (\ref{peakontrain-soln}).
The limiting procedure is described in detail in \cite{Ma07}. Due to (\ref{lambda})
one can write for the discrete eigenvalues $k_n=i\kappa_n$ \b 2\kappa_j=(1-\omega \lambda_j)^{1/2}=
1-\frac{1}{2}\omega \lambda_j+\ldots. \label{kappa-limit}\e

The solution (\ref{X=f/f}) depends explicitly on $\kappa_j$
(\ref{kappa-limit}) and the limit can be computed with
(\ref{kappa-limit}) by taking $\omega \to 0$ and keeping the
eigenvalue $\lambda_j$ constant. The result is the expression
(\ref{peakontrain-soln}) with \b p_i&=&{4D_{N-i+1}^{(0)}D_{N-i}^{(2)}\over D_{N-i+1}^{(1)}D_{N-i}^{(1)}},\qquad (i=1, 2, ..., N),\label{m_i}\\
q_i&=&\alpha {\text ln}\left[{2D_{N-i+1}^{(0)}\over D_{N-i}^{(2)}}\right],\qquad (i=1, 2, ..., N), \label{x_i} \e where \b D_n^{(m)}=\sum_{1\leq i_1<i_2<...<i_n\leq N}\Delta_n(i_1,i_2,
..., i_n)(\lambda_{i_1}\lambda_{i_2}...\lambda_{i_n})^m
R_{i_1}R_{i_2}...R_{i_n},\nonumber \\
  n=1, 2, ..., N, \nonumber \e
\b \Delta_n(i_1,i_2, ..., i_n)&=&\prod_{1\leq l<m\leq
n}(\lambda_{i_l}-\lambda_{i_m})^2,\qquad (n\geq 2), \\
R_i(t)&=&R_i(0)e^{\frac{2}{\lambda_i}t},\qquad(x_{i0}=\ln R_i(0)).\e

\noindent The quantities $D_n^{(m)}$ are called {\bfi Hankel determinants}. By definition $D_0^{(m)}=1$. In general, Hankel determinant is a determinant of $n\times n$ matrix of the form $D_n^{(m)}\equiv \det(a_{ij}^{(m)})$ where $a_{ij}$ are elements of a sequence, i.e. $$a_{ij}^{(m)}=A_{i+j+m-2}.$$ In this particular case \b A_l=\sum_{i=1}^{N}\lambda_i^l R_i(t). \e

\paragraph{Similarity of peakon lattice and Toda lattice}
Hankel determinants appear in the solutions of other integrable systems, e.g. Toda lattice, e.g. see \cite{GEI00}.
The Toda equation \cite{Toda} \b 
\frac{\text{d}p_n}{\text{d}t}&=&e^{q_{n+1}-q_n}-e^{q_n-q_{n-1}},
\nonumber \\
\frac{\text{d}q_n}{\text{d}t}&=&p_n, \quad n\in \mathbb{Z} \nonumber \e
is one of the most important integrable systems. We have a finite chain with
$N$ nodes under the fixed ends boundary conditions $q_0=-q_{N+1}=\infty$.
The solution is in the form \b  q_n(t)&=&q_1(0)+\ln\frac{D^{(0)}_n}{D^{(0)}_{n-1}}, \\
q_1(0)&=&-\frac{1}{N}\ln D^{(0)}_N(0) \qquad \text{is a constant}.\e
The Hankel determinants are obtained by a similar sequence, \b A_l&=&\sum_{i=1}^{N}\lambda_i^l R_i(t), \\
R_i(t)&=&R_i(0)e^{-\lambda_it},\e where $\lambda_i$ are $N$ different constants (eigenvalues of the Lax matrix)
and the quantities $R_i(0)$, $i=1,\ldots,N$ represent another set $N$ of constants.

Due to its simple form, the $N$-peakon solution can be used as an approximation
of the $N$-soliton CH solution when the dispersion term is small (and the
term $2\omega u_x$ can be neglected). Similarly, the Toda chain with complex
dynamical variables (the so-called Complex Toda Chain -- CTC) provides an approximation for the N-soliton solution of the Nonlinear Schr\"oedinger Equation (NLS) $$ iu_t+\frac{1}{2}u_{xx}+|u|^2 u =0, $$
see \cite{GKUE96,GEI98, GEI00} for more details. Such an approximation is called adiabatic approximation and means that the $N$-soliton solution
consists of $N $ well separated solitons 
$$ u(x,t)\approx \sum_{k=1}^{N}\frac{2\nu_k e^{i[2\mu_k (x-\xi_k(t)+\delta_k(t)]}}{\cosh
(2\nu_k(x-\xi_k(t))},$$ i.e. the overlap
between the solitons is small. The variables $q_n(t)$ of the CTC are related to the NLS solitons parameters by
$$
q_k (t)= - 2\nu_0  \xi_k(t)   +   i \left( 2\mu_0 \xi_k(t) - \delta_k(t)\right) + \text{const} $$
where $\xi_k $, $\delta _k $, $\mu_k=\frac{1}{2}\dot{\xi}_k$ and 
$\nu_k=(\frac{1}{2}\dot{\delta}_k- \mu_k^2)^{1/2}$ characterize the center-of-mass position, the phase, velocity and amplitude
respectively of the $k $-th soliton in the chain; $\nu _0 $ and $\mu _0 $ are the average amplitude and velocity of the soliton train. $\xi_k $ and $\delta _k $ can be obtained as the real and imaginary parts of $q_k(t)$.
Such soliton trains and their asymptotic behavior appear to be important for the needs of soliton based fiber optics communications.

\section{Superintegrability of the peakon system}\label{superint-sec}

Suppose we have an  integrable system with $2N$- dimensional phase space
which in terms of the Action-Angle (canonical) variables can be
represented as \b\dot{ \Lambda}_n=0,\qquad \dot{\Phi}_n=\Lambda_n, \qquad n=1,2, \ldots, N,\e

\noindent or, if there exists a bracket such that \footnote{The 1/2 coefficient
in the definition of the bracket appears in order to match it to the Poisson
bracket used in the CH Hamiltonian formulation.} \b \{\Phi_n, \Lambda_l\}
=\frac{1}{2}\delta_{nl},
\qquad \{\Phi_n, \Phi_l\}=\{\Lambda_n, \Lambda_l\}=0,\label{brackets}\e
the system is Hamiltonian with a Hamiltonian \b h_N=\Lambda_1^2+\ldots
\Lambda_N^2. \label{PHamilt}\e The integrals $\Lambda_n$, $n=1,2, \ldots, N$ are clearly in involution, which guarantees the integrability of the system. There is however another set of integrals:

\b I_j=(\Phi_j-\Phi_{j+1})(\Lambda_1+\ldots +\Lambda_N)-(\Lambda_j-\Lambda_{j+1})
(\Phi_1+\ldots\Phi_N)\nonumber
\\
\qquad j=1,2,\ldots,N-1.\label{I_j}\e

If the 'action' variables $\Lambda_n$ are all different, the set (\ref{I_j}) is functionally independent
from the set  $\Lambda_n$, $n=1,2, \ldots, N$. In addition, the integrals
(\ref{I_j}) form another set of $N$-integrals in involution together with $H$. Due to the existence of two sets of functionally independent integrals
in involution such systems are termed superintegrable. 

An example of such system is Toda lattice, see the discussion in \cite{PD06}.
The peakon system is also superintegrable. The canonical variables in terms
of the scattering data for CH equation can be used in the peakon limit:
$\Lambda_n = 2/\lambda_n$, $\Phi_n =\ln R_n(t)$, the Hamiltonian (\ref{PHamilt})
is also a peakon limit ($\omega \to 0$) of the $N$-soliton Hamiltonian \cite{CI06}
\b H_N(\omega) =
\omega^2\sum_{n=1}^{N}\Big(\ln\frac{1-2\kappa_n}{1+2\kappa_n}+
\frac{4\kappa_n(1+4\kappa_n^2)}{(1-4\kappa_n^2)^2}\Big).\nonumber
\e The Poisson bracket for the CH peakon solution is
\begin{equation}\label{PB}
\{A,B\}\equiv -\int_{-\infty}^{\infty} \frac{\delta A}{\delta m}(
m\partial+\partial m)\frac{\delta B}{\delta m}\text{d}x.
\end{equation} and the scattering data satisfy (\ref{brackets}) with respect
to (\ref{PB}), see \cite{CI06} for the details. 

 More interesting are the integrals (\ref{I_j}). From 
(\ref{x_i}) and (\ref{m_i}) one can recover these integrals in coordinate form. For example, when $N=2$ we have

\b I_1& =&\ln\frac{\sqrt{J}+p_1-p_2}{\sqrt{J}-p_1+p_2}+\frac{\sqrt{J}}{p_1+p_2}\left(\frac{q_1+q_2}{\alpha}+
\ln \frac{p_1}{p_2}\right), 
\\
&\text{where} &  J=(p_1-p_2)^2+4p_1p_2e^{-|q_1-q_2|/\alpha} . \e

Note that $I_1$ depends on both combinations $q_1+q_2$ and  $q_1-q_2$
as well as the momentum variables. The Hamiltonian $h_2$ (which depends only on $q_1-q_2$ and the momentum variables) and $I_1$ form a
complete system of integrals in involution. The integration of the 2-peakon
system with these integrals can be performed as follows. First, one can express
$q_1$ and $q_2$ in terms of $I_1$, $h_2$ and the momentum variables: $q_i=q_i(p_1,p_2,I_1,h_2)$.
Next, \b \dot{q}_i=\frac{\partial q_i}{\partial p_1}\dot{p_1}+\frac{\partial q_i}{\partial p_2}\dot{p_2}.\label{alg}\e The substitution of $\dot{q}_i$, $\dot{p}_i$ from (\ref{q_a}) (\ref{p_a}) to (\ref{alg}) produces an algebraic equation that gives, say $p_2$ as a function of $p_1$. Then (\ref{p_a}) is
an ODE for $p_1$ of the form $ \dot{p}_1=f(p_1,I_1,h_2)$.
  Clearly, from practical viewpoint
it is much more convenient to work with the other system of  integrals in involution: $h_2$ and the conserved momentum $P=p_1+p_2$. Note that $J=4h_2-3P^2$ is itself an integral.

\section{Other singular solutions: the
dispersionless $b$-equation} \label{other-sing-sol-sec}

\paragraph{Pulsons: Generalizing the peakon solutions of the
dispersionless $b$-equation for other Green's functions.}  The 
Hamiltonian $h_N$ in equation (\ref{H-peakon-ansatz}) depends on the Green's
function for the relation between velocity $u$ and momentum $m$. However, the
singular momentum solution ansatz (\ref{CHpeakon-m-soln}) is {\it independent}
of this Green's function. Thus, as discovered in Fringer and Holm
\cite{FrHo2001}, 

\noindent
{\it The singular momentum solution ansatz
(\ref{CHpeakon-m-soln}) for the dispersionless equation,
\begin{equation}\label{FHmom-eqn}
m_t+um_x+2mu_x=0,\quad \hbox{with}\quad u=g*m,
\end{equation}
provides an invariant manifold on which canonical Hamiltonian dynamics occurs,
for {\it any choice} of the Green's function $g$ relating velocity $u$ and
momentum $m$ by the convolution $u=g*m$. }

The fluid velocity solutions corresponding to the singular momentum ansatz  
(\ref{CHpeakon-m-soln}) for equation (\ref{FHmom-eqn}) are the {\it pulsons}.
Pulsons are given by the sum over $N$ velocity profiles determined by the
Green's function $g$, as
\begin{equation}\label{pulson-u-soln}
u(x,t)=\sum_{a=1}^N \, p_a(t)g\big(x,q_a(t)\big)
\,.
\end{equation}
Again for (\ref{FHmom-eqn}), the singular momentum ansatz
(\ref{CHpeakon-m-soln}) results in a finite-dimensional invariant manifold of
solutions, whose dynamics is canonically Hamiltonian.  The Hamiltonian for the
canonical dynamics of the $2N$ parameters $p_a(t)$ and
$q_a(t)$ in the ``pulson'' solutions (\ref{pulson-u-soln}) of equation
(\ref{FHmom-eqn}) is 
\begin{equation}\label{H-pulson-ansatz}
h_N=\tfrac{1}{2}\sum_{a,b=1}^N p_a\,p_b\,g(q_a,q_b)
\,.
\end{equation}
Again for the pulsons, the canonical equations for the invariant manifold of
singular momentum solutions provide a phase-space description of geodesic
motion, this time with respect to the co-metric given by the Green's function
$g$. Mathematical analysis and numerical results for the dynamics of these
pulson solutions are given in \cite{FrHo2001}. These results describe how the
collisions of  pulsons (\ref{pulson-u-soln}) depend upon their shape.%

\paragraph{Compactons in the $1/\alpha^2\to0$ limit of CH.} As mentioned
earlier, in the limit that $\alpha^2\to0$, the CH equation (\ref{CH-u-eqn})
becomes the KdV equation. In the opposite limit that $1/\alpha^2\to0$ CH
becomes the Hunter-Zheng equation \cite{HuZh1994}
\[
\big(u_t+uu_x\big)_{xx} = \frac{1}{2}(u_x^2)_x
\hspace{2cm}(\hbox{Hunter-Zheng})
\]
This equation has ``compacton'' solutions, whose collision dynamics was
studied numerically and put into the present context in \cite{FrHo2001}. The
corresponding Green's function satisfies $-\partial_x^2g(x)=2\delta(x)$, so it
has the triangular shape, $g(x)=1-|x|$ for $|x|<1$, and vanishes otherwise,
for $|x|\ge1$. That is, the Green's function in this case has compact support;
hence, the name ``compactons'' for these pulson solutions, which as a
limit of the integrable CH equations are true solitons, solvable by IST. 

\paragraph{Pulson solutions of the dispersionless $b$-equation.}
Holm and Staley \cite{HoSt2003a} give the pulson solutions of the
traveling wave problem and their elastic collision properties for
the dispersionless $b$-equation,
\begin{equation}\label{disp-b-eqn}
m_t
+
um_x
+
b\,mu_x
=0
\,,
\quad\hbox{with}\quad
u=g*m
\,,
\end{equation}
with any (symmetric) Green's function $g$ and for any value of the parameter
$b$. Numerically, pulsons and peakons are both found to be stable for $b>1$,
\cite{HoSt2003a}. The reduction to {\it noncanonical} Hamiltonian
dynamics for the invariant manifold of singular momentum solutions
(\ref{CHpeakon-m-soln}) of the other integrable case $b=3$ with peakon Green's
function $g(x,y)=e^{-|x-y|/\alpha}$ is found in
\cite{DePr1999}.

%%%%%%%%%%%%%%%%%%%%%%%%%%%%%%%%%%%%%%%%%%%%%%%%%%%%%%%%%%%%%%%%%%%%%%%%

\section{Euler-Poincar\'e theory in higher dimensions}\label{EP-higher-sec}

\paragraph{Generalizing the peakon solutions of the CH equation to
higher dimensions.}  
In \cite{HoSt2003a}, weakly nonlinear analysis and the assumption of
columnar motion in the variational principle for Euler's equations were found
to produce the two-dimensional generalization of the dispersionless CH
equation (\ref{DCHmom-eqn}). This generalization is the Euler-Poincar\'e (EP)
equation
\cite{HoMaRa1998a} for the Lagrangian consisting of the kinetic energy,
\begin{equation}
\ell%(\{{\mathbf{u}}\},\{D\})
=
\frac{1}{2}\int 
\Big[|{\mathbf{u}}|^2
+
\alpha^2\big(
{\rm div\,}{\mathbf{u}}
\big)^2
\Big]dxdy
\,,
\label{KE-def}
\end{equation} 
in which the fluid velocity ${\mathbf{u}}$ is a two-dimensional vector.
Evolution generated by kinetic energy in Hamilton's principle results in
geodesic motion, with respect to the velocity norm $\|{\mathbf{u}}\|$, which
is  provided by the kinetic energy Lagrangian. For ideal incompressible
fluids governed by Euler's equations, the importance of geodesic flow was
recognized by Arnold \cite{Ar1966} for the $L^2$ norm of the fluid velocity.
The EP equation generated by any choice of kinetic energy norm without
imposing incompressibility is called ``EPDiff,'' for ``Euler-Poincar\'e
equation for geodesic motion on the diffeomorphisms.''  EPDiff is given by
\cite{HoMaRa1998a}
\begin{equation}\label{EPDiff-eqn}
\Big(\frac{\partial}{\partial t}
+
{\mathbf{u}}\cdot\nabla\Big)
\mathbf{m}
+
\nabla \mathbf{u}^T\cdot
\mathbf{m}
+
\mathbf{m}
({\rm div\,}{\mathbf{u}})
=
0
\,,\end{equation}
with momentum density 
$\mathbf{m} 
= 
\delta\ell/\delta{\mathbf{u}}
\,,$
where $\ell=\frac{1}{2}\|{\mathbf{u}}\|^2$ is given by the kinetic energy,
which defines a norm in the fluid velocity $\|{\mathbf{u}}\|$, yet to be
determined. By design, this equation has no contribution from either potential
energy, or pressure. It conserves the velocity norm $\|{\mathbf{u}}\|$ given
by the kinetic energy. Its evolution describes  geodesic motion on the
diffeomorphisms with respect to this norm
\cite{HoMaRa1998a}. An alternative way of writing the EPDiff equation
(\ref{EPDiff-eqn}) in either two, or three dimensions is,
\begin{equation}\label{H1-EPcurl-eqn}
\frac{\partial}{\partial t}
\mathbf{m}
-
{\mathbf{u}}\times{\rm curl\,}{\mathbf{m}}
+
\nabla({\mathbf{u}}\cdot{\mathbf{m}})
+
\mathbf{m}
({\rm div\,}{\mathbf{u}})
=
0
\,.\end{equation}
This form of EPDiff involves all three differential operators, curl,
gradient and divergence. 
For the kinetic energy Lagrangian $\ell$ given in (\ref{KE-def}), which is a
norm for {\it irrotational} flow (with ${\rm curl\,}{\mathbf{u}}=0$), we have 
the EPDiff equation (\ref{EPDiff-eqn}) with momentum $\mathbf{m}  = 
\delta\ell/\delta{\mathbf{u}}
=
\mathbf{u}
-
\alpha^2\nabla({\rm div\,}{\mathbf{u}})$. 

EPDiff (\ref{EPDiff-eqn}) may also be written intrinsically as
\begin{equation}\label{EPDiffeo-eqn}
\frac{\partial}{\partial t}
\frac{\delta\ell}{\delta {\mathbf{u}}}
=
-
{\,\rm ad}^*_{\mathbf{u}}
\frac{\delta\ell}{\delta {\mathbf{u}}}
\,,
\end{equation}
where ad$^*$ is the $L^2$ dual of the ad-operation (commutator) for vector
fields. See \cite{ArKh1998,MaRa1999} for additional discussions of the
beautiful geometry underlying this equation.

\paragraph{Reduction to the dispersionless CH equation in 1D.}
In one dimension, the EPDiff equation (\ref{EPDiff-eqn}-\ref{EPDiffeo-eqn})
with Lagrangian
$\ell$ given in (\ref{KE-def}) simplifies to the dispersionless CH equation
(\ref{DCHmom-eqn}). The dispersionless limit of the CH equation
appears, because we have ignored potential energy and pressure.

\paragraph{Strengthening the kinetic energy norm to allow for circulation.}
The kinetic energy Lagrangian (\ref{KE-def}) is a norm for irrotational flow,
with ${\rm curl\,}{\mathbf{u}}=0$. However, inclusion of rotational
flow requires the kinetic energy norm to be strengthened to the $H_\alpha^1$
norm of the velocity, defined as
\begin{eqnarray}
\ell
&=&
\frac{1}{2}
\int 
\bigg[|{\mathbf{u}}|^2
+
\alpha^2\big(
{\rm div\,}{\mathbf{u}}
\big)^2
+
\alpha^2\big(
{\rm curl\,}{\mathbf{u}}
\big)^2
\bigg]dxdy
\nonumber\\
&=&
\frac{1}{2}
\int 
\bigg[|{\mathbf{u}}|^2
+
\alpha^2|\nabla{\mathbf{u}}|^2
\bigg]dxdy
=
\frac{1}{2}
\|{\mathbf{u}}\|^2_{H_\alpha^1}
\,.
\label{H1-eqn}
\end{eqnarray} 
Here we assume boundary conditions that give no contributions upon  
integrating by parts. The corresponding EPDiff equation is (\ref{EPDiff-eqn})
with $\mathbf{m}
\equiv
\delta \ell/\delta \mathbf{u}
=
\mathbf{u}
-
\alpha^2\Delta{\mathbf{u}}
\,.$
This expression involves inversion of the familiar {\it Helmholtz} operator in
the (nonlocal) relation between fluid velocity and momentum density. The
$H_\alpha^1$ norm $\|{\mathbf{u}}\|^2_{H_\alpha^1}$ for the kinetic energy
(\ref{H1-eqn}) also arises in three dimensions for turbulence modeling based
on Lagrangian averaging and using Taylor's hypothesis that the turbulent
fluctuations are ``frozen'' into the Lagrangian mean flow
\cite{FoHoTi2001}. 

\paragraph{Generalizing the CH peakon solutions to $n$ dimensions.}
Building on the peakon solutions (\ref{peakontrain-soln}) for the CH equation
and the pulsons (\ref{pulson-u-soln}) for its generalization to other
traveling-wave shapes in \cite{FrHo2001}, Holm and Staley
\cite{HoSt2003a} introduced the following measure-valued singular
momentum solution ansatz for the
$n-$dimensional solutions of the EPDiff
equation (\ref{EPDiff-eqn}):\footnote{These solutions represent smooth embeddings ${\rm Emb}(\mathbb{R}^k,\mathbb{R}^n)$ with $k<n$. In contrast, the similar expression (\ref{m(x,t)}) for the soliton solutions represent smooth functions $\mathbb{R}\to\mathbb{R}$.}
\begin{equation}\label{m-ansatz}
\mathbf{m}(\mathbf{x},t)
=
\sum_{a=1}^N\int\mathbf{P}^a(s,t)\,
\delta\big(\,\mathbf{x}-\mathbf{Q}^a(s,t)\,\big)ds.
\end{equation}
These singular momentum solutions, called ``diffeons,'' are vector density
functions supported in ${\mathbb{R}}^n$ on a set of $N$ surfaces (or curves)
of  codimension $(n-k)$ for $s\in {\mathbb{R}}^{k}$ with $k<n$.  They may, for
example, be supported on sets of points (vector peakons, $k=0$),
one-dimensional filaments (strings, $k=1$), or two-dimensional surfaces
(sheets, $k=2$) in three dimensions. 

%%%%%%%%%%%%%%%%%%%%%%%%%%%%%%%%%%%%%%%%%%%%%%%%%%%%%%%%%%%%%%%%%%%%%%%%%%%%

\begin{figure}[ht]
\centering
\includegraphics[scale=0.75,angle=0]{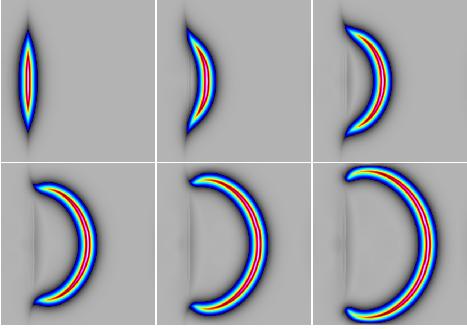}
\caption{%\footnotesize
A peakon segment of finite length is initially moving
rightward (East). Because its speed vanishes at its ends and it has fully 
two-dimensional spatial dependence, it expands into a peakon ``bubble''
as it propagates.  (The colors indicate speed: red is highest, yellow is
less, blue low, grey zero.) }
\label{2Dstrip_plate}
\end{figure}

Figure \ref{2Dstrip_plate}
shows the results for the EPDiff equation when a straight peakon segment of
finite length is created initially moving rightward (East). Because of  
propagation along the segment in adjusting to the condition of zero
speed at its ends and finite speed in its interior, the initially straight
segment expands outward as it propagates and curves into a peakon ``bubble.''

%%%%%%%%%%%%%%%%%%%%%%%%%%%%%%%%%%%%%%%%%%%%%%%%%%%%%%%%%%%%%%%%%%%%%%%%%%%%

\begin{figure}[ht]
\begin{center}
\includegraphics[scale=0.75,angle=0]{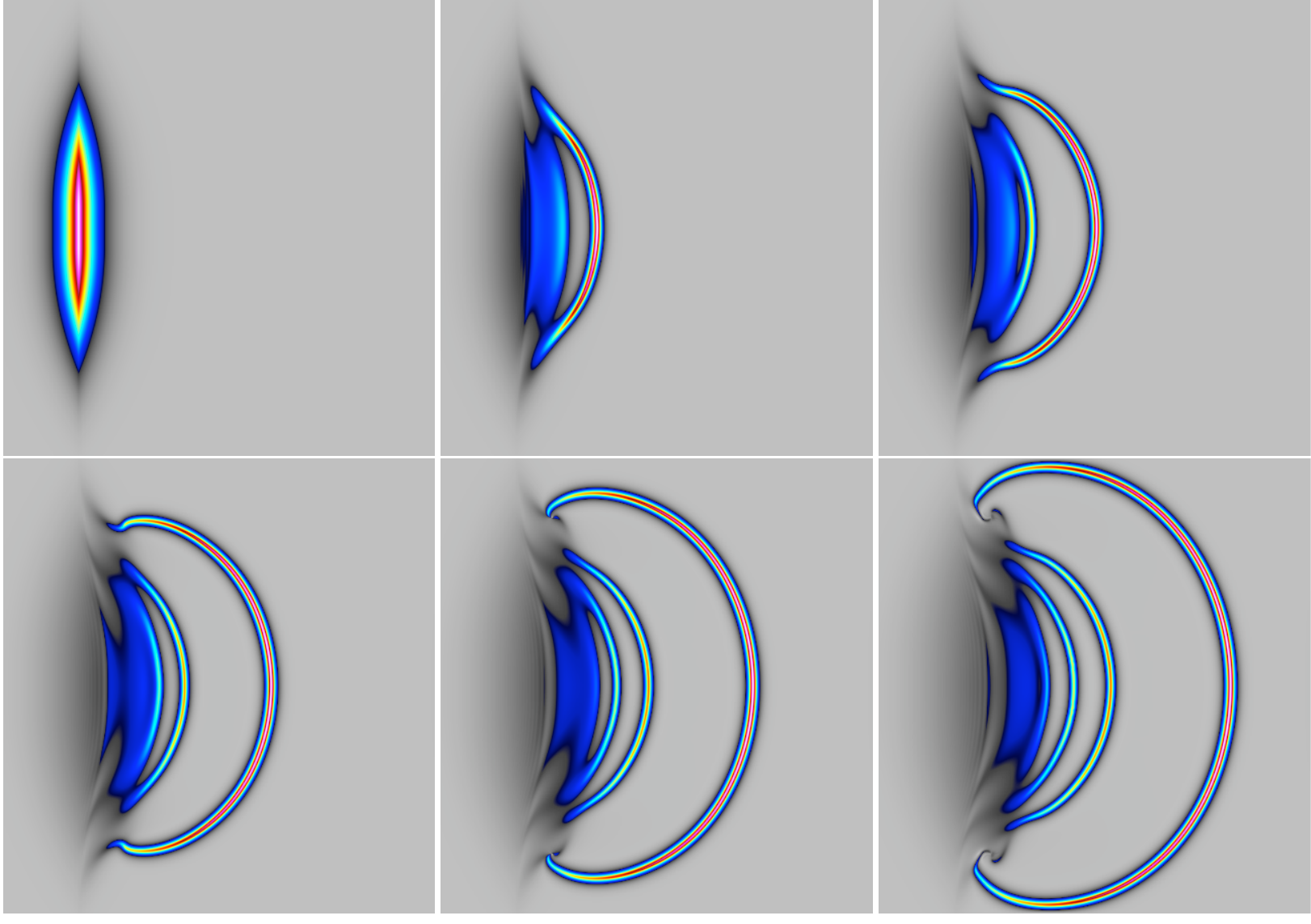}
\end{center}
\caption{%\footnotesize
An initially straight segment of velocity distribution whose exponential
profile is wider than the width $\alpha$ for the peakon solution will break
up into a train of curved peakon ``bubbles,'' each of width $\alpha$. This
example illustrates the emergent property of the peakon solutions in two
dimensions. }
\label{2Dstrip_plate-multi}
\end{figure}

Figure \ref{2Dstrip_plate-multi} shows an
initially straight segment whose velocity distribution is exponential in
the transverse direction, but is wider than $\alpha$ for the peakon solution.
This initial velocity distribution evolves under EPDiff to separate into a
train of curved peakon ``bubbles,'' each of width $\alpha$. This
example illustrates the emergent property of the peakon solutions in two
dimensions. This phenomenon is observed in nature, for example, as trains of
internal wave fronts in the south China Sea \cite{Liu-etal[1998]}.

%%%%%%%%%%%%%%%%%%%%%%%%%%%%%%%%%%%%%%%%%%%%%%%%%%%%%%%%%%%%%%%%%%%%%%%%%%%%

%%%%%%%%%%%%%%%%%%%%%%%%%%%%%%%%%%%%%%%%%%%%%%%%%%%%%%%%%%%%%%%%%%%%%%%%%%%%
\begin{figure}[ht]
\begin{center}
\includegraphics[scale=0.75,angle=0]{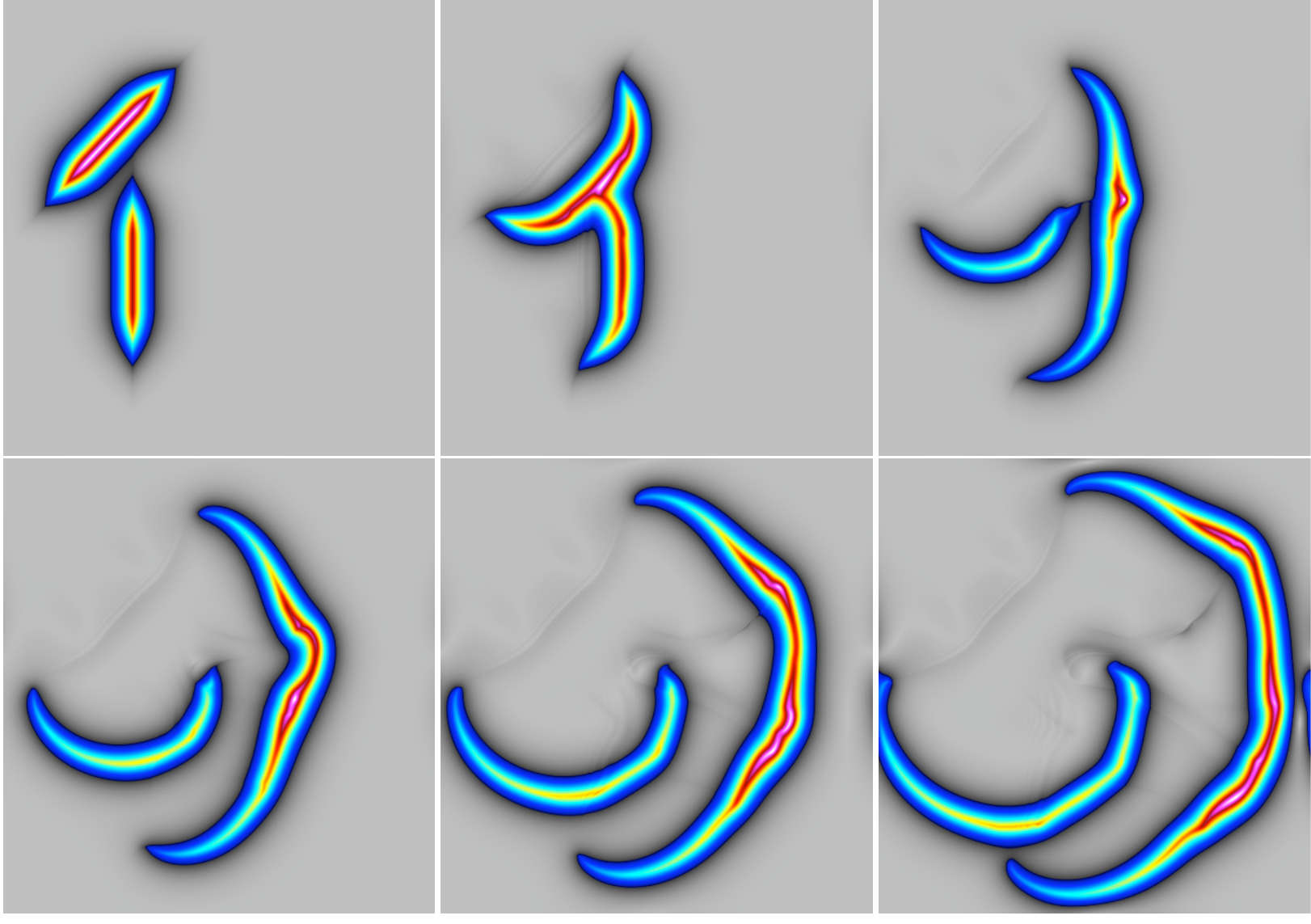}
\end{center}
\caption{%\footnotesize
A single collision is shown involving reconnection as the faster
peakon segment initially moving Southeast along the diagonal expands, curves 
and obliquely overtakes the slower peakon segment initially moving rightward
(East). This reconnection illustrates one of the collision rules for the
strongly two-dimensional EPDiff flow. }
\label{2Dstrip_skew}
\end{figure}

 Substitution of the singular momentum solution ansatz
(\ref{m-ansatz}) into the EPDiff
equation (\ref{EPDiff-eqn}) implies the following 
integro-partial-differential equations (IPDEs) for the evolution of
the parameters $\{{\mathbf{P}}\}$ and $\{{\mathbf{Q}}\}$,
%%%%%%%%%%%%%%%%%%%%%%%%%%%%%%%%%%%%%%%%%%%%%%%%%%%%%%%%%%%%%%%%%%%%
\begin{eqnarray}
%\hspace{-3mm}
\frac{\partial }{\partial t}\mathbf{Q}^a (s,t)
&=&
\sum_{b=1}^{N} \int\mathbf{P}^b(s^{\prime},t)\,
G(\mathbf{Q}^a(s,t)-\mathbf{Q}^b(s^{\prime},t)\,\big)ds^{\prime}
\,,\label{IntDiffEqn-Q}
\nonumber\\
%\hspace{-3mm}
\frac{\partial }{\partial t}\mathbf{P}^a (s,t)
&=&
-\,\!\!\sum_{b=1}^{N} \int
\big(\mathbf{P}^a(s,t)\!\cdot\!\mathbf{P}^b(s^{\prime},t)\big)
\\&&\hspace{5mm}
\frac{\partial }{\partial \mathbf{Q}^a(s,t)}
G\big(\mathbf{Q}^a(s,t)-\mathbf{Q}^b(s^{\prime},t)\big)\,ds^{\prime}
\,.
\nonumber
\end{eqnarray}
%%%%%%%%%%%%%%%%%%%%%%%%%%%%%%%%%%%%%%%%%%%%%%%%%%%%%%%%%%%%%%%%%%%%%%%%
%
Importantly for the interpretation of these solutions, the coordinates
$s\in {\mathbb{R}}^{k}$ turn out to be {\it Lagrangian} coordinates. The 
velocity field corresponding to the momentum solution ansatz
(\ref{m-ansatz}) is given by
\begin{eqnarray}\label{u-ansatz}
\mathbf{u}(\mathbf{x},t)
&=&
G*\mathbf{m}
\nonumber\\
&=&
\sum_{b=1}^N\int\mathbf{P}^b(s^{\prime},t)\,
G\big(\,\mathbf{x}-\mathbf{Q}^b(s^{\prime},t)\,\big)ds^{\prime}
\,,
\end{eqnarray}
for $\mathbf{u}\in{\mathbb{R}}^n$.
When evaluated along the curve $\mathbf{x}=\mathbf{Q}^a(s,t)$, this
velocity satisfies,
\begin{eqnarray}\label{Qdot-ansatz}
\mathbf{u}(\mathbf{Q}^a(s,t),t)
&=&
\sum_{b=1}^N\int\mathbf{P}^b(s^{\prime},t)\,
G\big(\,\mathbf{Q}^a(s,t)
-\mathbf{Q}^b(s^{\prime},t)\,\big)ds^{\prime}
\nonumber \\
&=&
\frac{\partial\mathbf{Q}^a(s,t)}{\partial t}
\,.
\end{eqnarray}
Consequently, the lower-dimensional support sets defined on
$\mathbf{x}=\mathbf{Q}^a(s,t)$ and parameterized by coordinates
$s\in{\mathbb{R}}^{k}$ move with the fluid velocity. This means the
$s\in{\mathbb{R}}^{k}$ are Lagrangian coordinates. Moreover, equations
(\ref{IntDiffEqn-Q}) for the evolution of these support sets are canonical
Hamiltonian equations,
%-----------------------------
\begin{equation}
\label{IntDiffEqns-Ham}
\frac{\partial }{\partial t}\mathbf{{Q}}^a (s,t)
=
\frac{\delta H_N}{\delta \mathbf{P}^a}
\,,\qquad
\frac{\partial }{\partial t}\mathbf{{P}}^a (s,t)
=
-\,\frac{\delta H_N}{\delta \mathbf{Q}^a}
\,.
\end{equation}
%-----------------------------
The corresponding Hamiltonian function 
$H_N:({\mathbb{R}}^n\times {\mathbb{R}}^n)^{N}\to {\mathbb{R}}$ is,
%-----------------------------
\begin{eqnarray} 
H_N &=& \frac{1}{2}\!\int\!\!\!\!\int\!\!\sum_{a\,,\,b=1}^{N}
\big(\mathbf{P}^a(s,t)\cdot\mathbf{P}^b(s^{\prime},t)\big)
\nonumber\\&&\hspace{1cm}
G\big(\mathbf{Q}^a(s,t),\mathbf{Q}^b(s^{\prime},t)\big)
\,ds\,ds^{\prime}
\,.\label{H_N-def}
\end{eqnarray}
%-----------------------------
This is the Hamiltonian for geodesic motion on the cotangent bundle of
a set of curves $\mathbf{Q}^a(s,t)$ with respect to the metric given by
$G$. This dynamics was investigated numerically in 
\cite{HoSt2003a} to which we refer for more details of
the solution properties. One important result found {\it numerically} in 
\cite{HoSt2003a} is that only codimension-one singular momentum solutions 
appear to be stable under the evolution of the EPDiff equation. Thus, we have

{\it Stability for codimension-one: the singular momentum solutions of
EPDiff are stable, as points on the line (peakons), as curves in the plane
(filaments, or wave fronts), or as surfaces in space (sheets). }

Proving this stability result analytically remains an outstanding problem. The
stability of peakons on the real line is proven in \cite{CoSt2000}.

%%%%%%%%%%%%%%%%%%%%%%%%%%%%%%%%%%%%%%%%%%%%%%
\rem{
%%%%%%%%%%%%%%%%%%%%%%%%%%%%%%
\begin{figure}[ht]
\begin{center}
\includegraphics[scale=0.75,angle=0]{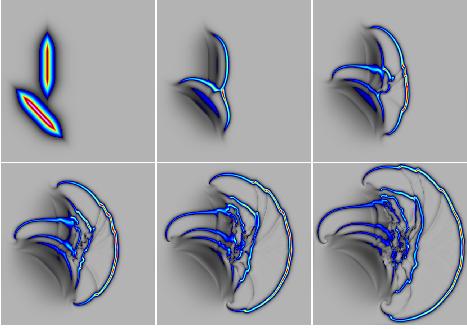}
\end{center}
\caption{%\footnotesize
A series of multiple collisions is shown involving reconnections as the
faster wider peakon segment initially moving Northeast along the diagonal
expands, breaks up into a wave train of peakons, each of which propagates,
curves and obliquely overtakes the slower wide peakon segment initially moving
rightward (East), which is also breaking up into a train of wavefronts. In
this series of oblique collision, the now-curved peakon filaments exchange
momentum and reconnect several times. }
\label{2Dstrip_skew-multi}
\end{figure}
}
%%%%%%%%%%%%%%%%%%%%%%%%%%%%%%%%%%%%%%%%%%%%%%%%%%%%%%%%%%%%%%%%%%%%%%%%%%%%
\paragraph{Reconnections in oblique overtaking collisions of peakon wave fronts.}  
Figures \ref{2Dstrip_skew} shows 
results of oblique wave front collisions producing reconnections for the
EPDiff equation in two dimensions. Figure \ref{2Dstrip_skew} shows a single
oblique overtaking collision, as a faster expanding peakon wave front
overtakes a slower one and reconnects with it at the collision point. 
\rem{
Figure
\ref{2Dstrip_skew-multi} shows a series of reconnections involving the
oblique overtaking collisions of two trains of curved peakon filaments, or
wave fronts.
}

%%%%%%%%%%%%%%%%%%%%%%%%%%%%%%%%%%%%%%%%%%%%%%

%%%%%%%%%%%%%%%%%%%%%%%%%%%%%%
\paragraph{The peakon solution ansatz is a momentum map.} As shown in \cite{HoMa2004}, the solution expressions (\ref{CHpeakon-m-soln}) in one dimension and (\ref{m-ansatz}) in higher dimensions may be interpreted as equivariant momentum maps, from the cotangent bundle of the smooth embeddings of lower dimensional
sets ${\mathbb{R}}^s\subset{\mathbb{R}}^n\,,$ to the dual of the Lie algebra of
vector fields defined on these sets. (Momentum maps for Hamiltonian
dynamics are reviewed in \cite{MaRa1999}, for example.) 
The result that the singular solution ansatz
\textup{(\ref{m-ansatz})} is a momentum map helps to organize the theory, to
explain previous results and to suggest new avenues of exploration.
This geometric feature underlies the remarkable reduction properties of the CH and EPDiff equations, and \emph{explains} why they must be Lie-Poisson Hamiltonian equations. This is because of the general fact that equivariant momentum maps are Poisson maps. This geometric feature also underlies the singular momentum
solution (\ref{m-ansatz}) and its associated velocity (\ref{u-ansatz})
which generalize the peakon solutions, both to higher dimensions and to
arbitrary kinetic energy metrics.  As we saw in Section \ref{mommap-sec}, the soliton solution (\ref{mommap1}) is also  a momentum map. This soliton momentum map may be expected to apply in the action-angle representation of the solution of any integrable Hamiltonian PDE. Its further properties will be studied in detail elsewhere.

\section{Three open problems}\label{Conclus-sec}

(1.) Throughout this discussion the solutions $u(x,t)$ were confined to be functions in the Schwartz class, $\omega>0$.  The situation when the condition
$m(x,0)+\omega>0$ on the initial data does not hold is more complicated
and requires separate analysis \cite{K05,C01,CE98}. In general, it
leads to wave-breaking \cite{CE98}. An attempt at developing the inverse scattering theory for this case has been made by Kaup \cite{K05}, who suggested applying the inverse scattering approach separately in each interval where $m(x,t)+\omega$ is of the same sign. The problem however is how to join solutions that are valid in different intervals.

(2.) The peakon solution (\ref{m_i}) was obtained from the soliton solution under the assumption $m(x,0)+\omega>0$. Thus, all $p_k$ are of the same sign, since all the eigenvalues $\lambda_n$ in this case are positive. However,
one can formally use the same solution with
eigenvalues of various signs to model $p_k$ of various signs (mixture of peakons and `anti-peakons') and thus to study peakon --  anti-peakon interactions, see e.g. \cite{P08}.
The result is that the multi-peakon interaction (including anti-peakons) in general  decomposes into a sequence of pairwise collisions \cite{FrHo2001}. The collision of a single peakon-anti-peakon pair was studied already in \cite{CaHo1993}.
When the eigenvalues are of mixed signs, the Hankel determinants in the denominator of (\ref{m_i}) may develop singularities for finite
values of $t$. This `peakon-breaking' phenomenon is apparently the analog of the wave-breaking mentioned earlier when $\omega \rightarrow 0$. This needs to be investigated further. 

(3.) Stability for EPDiff singular momentum solutions, that is, proving their stability analytically, remains an outstanding problem.

%%%%%%%%%%%%%%%%%%%%%%%%%%%%%%%%%%%%%%%%%%%%%%%%%%%%%%%%%%%%%%%
\paragraph{Acknowledgments.} DDH is grateful to R. Camassa, J. E. Marsden, T. S. Ratiu and A. Weinstein for their collaboration, help and inspiring discussions over the years. DDH gratefully acknowledges partial support by the Royal Society of London's Wolfson Award scheme. RII acknowledges funding from a Marie Curie Intra-European Fellowship. 
Both authors thank J. Percival and M. F. Staley for providing figures.

\end{document}